\documentclass[final]{siamltex1213}
\usepackage[utf8]{inputenc}
\usepackage{amsmath}
\usepackage{graphicx}
\usepackage{hyperref}

\newcommand{\Pe}{\,\text{Pe}\,}
\newcommand{\Pep}{\text{Pe}^{\prime}}
\newcommand{\erfi}{\text{erfi}}

\newcommand{\Ei}{\text{Ei}}

\title{STEADILY TRANSLATING PARABOLIC DISSOLUTION FINGERS}
\author{PAWE{\L} KONDRATIUK\footnotemark[2] \and PIOTR SZYMCZAK\footnotemark[2]}

\begin{document}
\maketitle

\renewcommand{\thefootnote}{\fnsymbol{footnote}}
\footnotetext[2]{Institute of Theoretical Physics, Faculty of Physics, University of Warsaw, Pasteura 5, 02-093 Warsaw (pawel.kondratiuk@fuw.edu.pl, piotr.szymczak@fuw.edu.pl)}
\renewcommand{\thefootnote}{\arabic{footnote}}

\begin{abstract}
Dissolution fingers (or wormholes) are formed during the dissolution of a porous rock as a result of nonlinear feedbacks between the flow, transport and chemical reactions at pore surfaces. We analyze the shapes and growth velocities of such fingers within the thin-front approximation, in which the reaction is assumed to take place instantaneously with the reactants fully consumed at the dissolution front. We concentrate on the case when the main flow is driven by the constant pressure gradient far from the finger, and the permeability contrast between the inside and the outside of the finger is finite. Using Ivantsov ansatz and conformal transformations we find the family of steadily translating fingers characterized by a parabolic shape. We derive the reactant concentration field and the pressure field inside and outside of the fingers and show that the flow within them is uniform. The advancement velocity of the finger is shown to be inversely proportional to its radius of curvature in the small P\'{e}clet number limit and constant for large P\'{e}clet numbers.
\end{abstract}

\begin{keywords}
free boundary problems, dissolution, convection-diffusion-reaction, porous media
\end{keywords}

\begin{AMS}
35R35, 86A60, 76S05, 35Q35, 35Q86, 30E25 
\end{AMS}

\section{Introduction}

Chemical erosion of a porous medium by a reactive fluid is relevant for many natural and industrial applications and is a crucial component in a number of geological pattern formation processes \cite{Meakin2010,Jamtveit1999}. Networks of caves and sinkholes are formed by the dissolution of limestone by CO$_2$-enriched water in karst areas \cite{Palmer1991,Szymczak2011}, the ascending magma dissolves the peridotite rocks, leading to formation of porous channels \cite{Aharonov1995}, matrix acidizing is used by petroleum engineers to enlarge the natural pores of the reservoirs \cite{Rowan1959} - in all of these processes the interplay of flow and reaction in an evolving geometry results in spontaneous formation of intricate patterns. Reactive flow systems with strong, nonlinear coupling between the transport and geometry evolution have been investigated in a number of experimental \cite{Daccord1987,Daccord1987a,Huang2015} and theoretical \cite{Pawell1996,Golfier2004,Dreybrodt1990,Kalia2007} studies, the latter often combined with numerical simulations.

\begin{figure}
\centering
\includegraphics[width=0.8\textwidth]{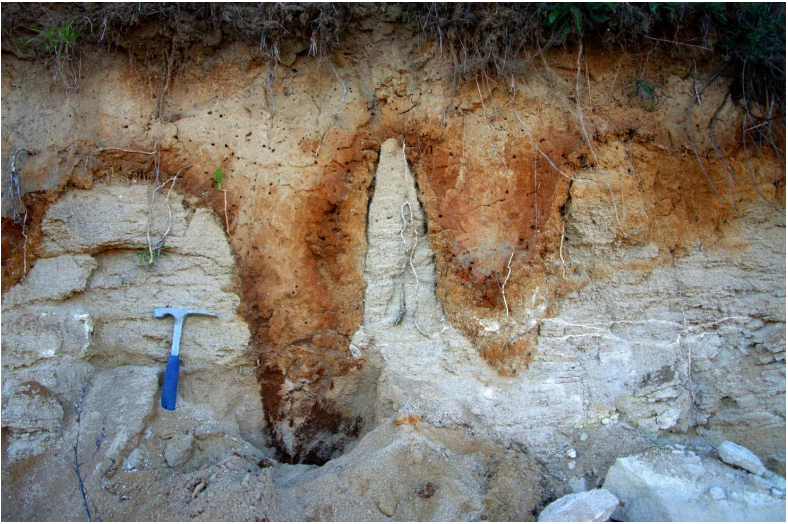}
\caption{Solution pipes in the limestone quarry in Smerdyna (Poland), formed in limestone bedrock. They are postulated to have been formed by a dissolving action of meltwater during Elsterian deglaciation  \cite{Morawiecka1997,Walsh2001}. The brown rims in the clay filling the pipes are formed as a result of clay and iron oxide accumulation due to the illuviation processes \cite{schaetzl2005}. The clay particles are transferred by water from the upper parts of the soil and then flocculated 
at the clay-limestone boundary, where the pH changes from mildly acidic to alkaline.}
\label{fig:solution_pipes}
\end{figure}

A locally increased dissolution rate can make the rock  more porous and thus more permeable. This increases the flow and reactant transport, further enhancing the local porosity increase. Such a positive feedback eventually leads to the creation of highly localized flow paths, which go by different names depending on the field. In petroleum industry, they are dubbed ``wormholes" \cite{Hoefner1988} due to the resemblance with the tunnels dug by worms, geologists call them ``solution pipes", ``karst funnels" or ``geological organs'' \cite{Jennings1985,Walsh2001,deWaele2011}, whereas pedologists - ``soil tongues'' \cite{yehle1954}. Examples of such structures exposed in the limestone quarry in Smerdyna, Poland, are presented in Fig. \ref{fig:solution_pipes}.

The initial stages of the wormhole development due to the breakup of the initially planar reaction front (referred to in the literature as the ``reactive-infiltration instability'') have been extensively studied \cite{Chadam1986,Ortoleva1987,Sherwood1987,Chadam1988,Hinch1990,Szymczak2014} and are now quite well understood. Much less is known, however, about the nonlinear regime, when the initial perturbations of the interface are transformed into finger-like structures that advance into the system \cite{Daccord1987,Golfier2002,Panga2005,Cohen2008,Szymczak2009}. An important question in this context concerns the existence of a self-preserving form that might model the growing tip. Such invariantly propagating forms have been found in other pattern forming systems, e.g. the Saffman-Taylor finger in viscous fingering or Ivantsov paraboloid in the dendritic growth \cite{Saffman1958,Horvay1961,Ivantsov1947,AddaBedia1994,Cummings1999}.  Our goal in this paper is to find such solutions for a dissolving porous medium.

In the context of dissolution, the propagation of individual fingers was studied in the petroleum engineering~\cite{Hung1989,Buijse2000,Cohen2008}. In these works, simple geometric models of wormhole shapes were adopted, which are not preserved during the growth. The closest in spirit to the present study is the work by Nilson and Griffiths \cite{Nilson1990}. They tackle the problem of steadily propagating dissolution forms and find them to be parabolic (in 2d) or paraboloidal (in 3d). However, there are significant differences between their approach and the present one. We discuss those in more detail at in Sec.~\ref{Nilson}. Here we will just note that in \cite{Nilson1990} the reactant concentration field is not resolved, instead the dissolution front velocity is taken to be proportional to the local fluid velocity. 

Additionally, the pressure drop inside the finger is neglected, which is only justified if the permeability of the dissolved phase is much higher than the permeability of the primary rock. In that way, the problem is effectively been reduced to a one-phase problem, requiring the solution of the Laplace equation for pressure in the region outside the finger only. As discussed in~\cite{Howison2000,Crowdy2006} such ``one-phase'' growth problems are in general much easier to solve than the ``two-phase'' problems in which the flow fields both inside and outside the finger need to be found, which is also the case in the present study. On top of it, to find the finger advancement velocity, we need to solve not only for the pressure, but also for the reactant concentration field, which makes the task at hand more challenging. 

The paper is organized as follows. In \S\ref{sec:model} the general equations governing the dynamics of the dissolving porous rock are briefly recalled and then put in the dimensionless form in \S\ref{sec:scaling}. The core of the paper are sections \ref{sec:2d} and \ref{sec:3d}, where - after the application of the Ivantsov ansatz - we obtain the family of steadily translating dissolution fingers and show that they are parabolic/paraboloidal. Let us reiterate that despite a formal similarity to the Ivantsov forms, the physics of growth is quite different here, as it involves an interplay between the flow field and solute concentration field in contrast to the dendritic growth which is controlled by a single field (temperature only).

\section{The model of matrix dissolution}
\label{sec:model}

\begin{figure}
\centering
\includegraphics[width=.8\textwidth]{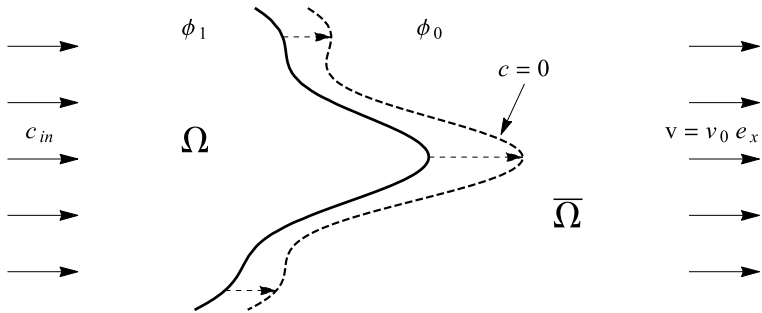}
\caption{Geometry of the system: a reactive fluid is injected from the left side and dissolves the porous matrix through chemical reactions. Once the dissolution is complete, the reaction front, shown by the solid
line, advances into the matrix (dashed line), separating the dissolved, upstream domain ($\Omega$) of porosity $\phi_1$ and the undissolved, downstream domain ($\bar{\Omega}$) of porosity $\phi_0$, complementary to $\Omega$.}
\label{fig:instab}
\end{figure}

When a porous matrix is infiltrated by an incoming flux of reactive fluid, a front develops once all the soluble material at the inlet has been dissolved. This front propagates into the matrix as illustrated in Fig.~\ref{fig:instab}. Upstream of the front, all the soluble material has dissolved and the porosity is constant, $\phi = \phi_1$. Ahead of the front the porosity decays gradually to its value in the undissolved matrix, $\phi = \phi_0$. The front is initially planar but eventually breaks up because of a positive feedback between flow and dissolution, which amplifies any small variation in the porosity field~\cite{Chadam1986}.

Let us briefly recall the equations for the dissolution of a porous matrix.  Rate of groundwater flow through the porous medium is taken to be proportional to the pressure gradient (Darcy's law),
\begin{equation}
\mathbf{v} = - K\frac{\nabla p}{\mu},
\label{e1}
\end{equation}
where $\mathbf{v} $ is the Darcy velocity, $\phi$ is the porosity and $K(\phi)$ is the permeability. We also assume that the Darcy velocity field is incompressible,
\begin{equation}\label{eq:vic}
\nabla \cdot \mathbf{v}  = 0,
\end{equation}
neglecting contributions to the fluid volume from reactants or dissolved products. Under typical geophysical conditions, dissolution is slow in comparison to flow and transport processes; we can therefore assume a steady state in both the flow and transport equations.

The transport of reactants is described by a convection-diffusion-reaction equation,
\begin{equation}\label{eq:CD}
\nabla \cdot (\mathbf{v}  c) - \nabla \cdot (D \phi \cdot \nabla c) = -  R,
\end{equation}
where $c$ denotes the concentration of the reactant and $R(c)$ is the reactive flux into the matrix. In the upstream region, where all the soluble material has dissolved ($\phi=\phi_1$), the reaction term vanishes and the transport equation reduces to a convection-diffusion equation.

In the derivations below, we adopt a thin-reaction-front approximation, in which the transition zone over which the porosity changes is assumed to be infinitely thin \cite{Chadam1986,Ortoleva1987}. As shown in \cite{Szymczak2013}, this approximation is applicable whenever the reactant penetration length is short in comparison with the diffusive length scale $D/v$, characterizing the decay of the concentration in the upstream region. In practice, this happens whenever the parameter $H=Dk/v^2 \gg 1$ \cite{Szymczak2013}, where $k$ is the dissolution rate (assuming the 1st order reaction kinetics); thus the reaction rate needs to be sufficiently high and/or the flow rate --- sufficiently low. As a consequence, we obtain a Stefan-like problem in which the space is divided into two domains: the dissolved, upstream domain ($\Omega$) of porosity $\phi_1$ and the undissolved, downstream domain ($\bar{\Omega}$) of porosity $\phi_0$, complementary to $\Omega$. The reaction front ($\partial \Omega$) advances with velocity proportional to the flux of the reactant at a given point
\begin{equation}
U_n = - \frac{\gamma}{c_{in}} D \left( \nabla c \right)_n 
\label{un}
\end{equation}
where subscript $n$ represents the component normal to the interface, $c_{in}$ is the inlet concentration of the reactant and the acid capacity number $\gamma = c_{in}/\nu c_{sol}(\phi_1-\phi_0)$ is defined as the volume of rock (of molar concentration $c_{sol}$) that is completely dissolved by a unit volume of reactant (of molar concentration $c_{in}$). 
Finally, $\nu$ is the stoichiometric coefficient in the dissolution reaction (number of moles of the reactant necessary to dissolve one mole of the rock).

The flux of reactant in \eqref{un} is taken to be purely diffusive, since the thin-front approximation implies vanishing of $c$ at the interface, as it is fully consumed there:
\begin{equation}
c\vert_{\partial \Omega(t)}= 0.
\end{equation}

Because in both domains the porosity (and hence permeability) is constant, Eqs.~(\ref{e1}--\ref{eq:vic}) can be combined to yield the Laplace equation for the pressure 
\begin{equation}
\nabla^2 p= 0
\end{equation}
in both $\Omega$ and $\bar{\Omega}$.

The above equations are supplemented by boundary conditions on the velocity and concentration field as $x \rightarrow \infty$:
\begin{align}
{\bf v}(x \rightarrow \infty) &=  v_0 {\bf e}_x, &  c(x\rightarrow\infty) = 0, \label{eq:vbc}
\end{align}
which accounts for the fact that far from the dissolution front the flow becomes uniform and the reactant concentration vanishes. The conditions at $x \rightarrow -\infty$ are somewhat more subtle. If the reaction front assumes such a form that sufficiently far upstream everything is dissolved, i.e. $\exists x_0 \left( \forall x<x_0, (x,y) \in \Omega \right)$, then it is possible to impose 
\begin{equation}
\partial_x{v_x}(x\rightarrow -\infty) = 0 
\label{flo}
\end{equation}
on the flow field and
\begin{equation}
c(x\rightarrow -\infty) = c_{in}, \label{eq:vbc2}
\end{equation}
on the concentration field; the former representing the condition that the flow becomes uniform far upstream, the latter corresponding to the reactant concentration imposed at the inlet. 

However, if the undissolved phase ($\bar{\Omega}$) extends towards $x \rightarrow -\infty$ (as it is the case for a solitary dissolution finger surrounded by an undissolved matrix), then the condition analogous to \eqref{eq:vbc2} should be imposed only {\it within} the finger, and not on the entire $x \rightarrow -\infty$ line. We will come back to this issue later on.

Additionally, both the pressure and the normal component of the Darcy velocity need to be continuous across the reaction front $\partial\Omega$:
\begin{align}
p \vert_{\partial \Omega(t)^-}&= p\vert_{\partial \Omega(t)^+}, \label{eq:p} \\
v_n \vert_{\partial \Omega(t)^-}&= v_n\vert_{\partial \Omega(t)^+} \label{eq:pp}
\end{align}

\section{Scaling of the variables and the limiting cases}\label{sec:scaling}

The dissolution equations (\ref{e1}--\ref{eq:CD}) can be simplified by scaling the velocity, concentration, and porosity fields by their characteristic values
\begin{equation}
\hat{\bf v}  = {\bf v}/v_0, \ \ \ \ \hat{c} = c/c_{in}, \ \ \ \ \hat{\phi} = \frac{\phi - \phi_0}{\phi_{1} - \phi_0},
\end{equation}
where the scaled, dimensionless variables are marked by hats. Additionally, we scale the spatial coordinates by some length $l$, characterizing the wormhole, and time by $\tau=l/\gamma v_0$. Using $l$, we can scale the pressure as follows
\begin{equation}
\hat p = \frac{K(\phi_0)}{v_0\mu l} p.
\end{equation}
With these scalings, the governing equations take the form 
\begin{align}
& \hat\nabla^2 \hat p = 0 &\quad {\bf r} \in \Omega(\hat t) \label{b0} \\
& \kappa \hat\nabla \hat p \cdot \hat\nabla \hat c + \Pe^{-1}\hat\nabla^2 \hat c = 0 &\quad {\bf r} \in \Omega(\hat t) \label{b1}
\end{align}
in the upstream region and
\begin{align}
& \hat\nabla^2 \hat p = 0 &\quad {\bf r} \in \bar{\Omega}(\hat t)\\
& \hat c = 0 &\quad {\bf r} \in \bar{\Omega}(\hat t) \label{b2}
\end{align}
in the downstream one. In the above,
\begin{equation}
\Pe \equiv \frac{v_0 l}{\phi_1 D(\phi_1)},
\end{equation}
is the P\'eclet number which measures the relative magnitude of convective and diffusive effects on the length scale $l$, whereas 
\begin{equation}
\kappa \equiv \frac{K(\phi_1)}{K(\phi_0)}.
\end{equation}
is the ratio of permeability between the domains. In typical geological conditions the permeability is an increasing function of porosity, thus  $\kappa>1$. The boundary conditions (\ref{eq:vbc}--\ref{eq:vbc2}) in the scaled variables take the form
\begin{align}
{\bf \hat{v}}(x \rightarrow \infty) &=  \hat{\bf e}_x, &  {\hat{c}}(x \rightarrow \infty) = 0, \label{eq:vbcc}
\end{align}
and
\begin{align}
(\partial_x \hat v_x)(x\rightarrow -\infty) &= 0, &  {\hat{c}}(x \rightarrow -\infty) = 1, \label{eq:vbc3}
\end{align}
which, as before, need to be supplemented by the continuity conditions for the pressure $\hat p$ and the normal component of the velocity $\hat v_n$ across the interface\footnote{Note that $\hat {\bf v} = -\kappa\hat\nabla \hat p$ on $\Omega$ and $\hat {\bf v} = -\hat\nabla \hat p$ on $\bar{\Omega}$.} at the boundary (reaction front) $\partial\Omega$.  
Finally, the condition \eqref{un} for the front advancement velocity takes the form
\begin{equation}
\hat{U}_n =  \Pe^{-1} (\hat \nabla \hat{c})_n
\label{un2}
\end{equation}
Noting that at the reaction front the concentration satisfies
\begin{equation}
\hat {\bf U} \cdot \hat \nabla \hat c + \frac{\partial \hat c}{\partial \hat t} = 0,
\end{equation}
one can rewrite \eqref{un2} in terms of $\hat c$ only:
\begin{equation}
 \frac{\partial\hat c (\hat{\bf r})}{\partial\hat t} -\Pe^{-1} |\hat\nabla \hat c  (\hat{\bf r})|^2 = 0 \quad  \ \ \ \ \ \ \ \ {\bf r} \in \partial\Omega(\hat t) \label{seq:bnd_eq}
\end{equation}

It is relatively straightforward to derive the one-dimensional solutions of Eqs.~(\ref{b0}--\ref{b2}), corresponding to the planar reactive front propagating with a constant velocity, $U_0$. Assuming that both $\hat c$ and $\hat p$ are the functions of $\hat x$ only, leads to 
\begin{equation}
\hat p(\tilde{x}) = \left\{ \begin{array}{l l}
-\kappa^{-1} \tilde{x} & \quad   \tilde{x} < 0 \\
-\tilde{x}  & \quad \tilde{x} > 0,
\end{array} \right.
\end{equation}
where $\tilde{x} = \hat x - \hat{U}_0 \hat t$ is the coordinate moving with the front. For the concentration we get an exponentially decaying, diffusive profile
\begin{equation}
\hat c(\tilde x) = \left\{ \begin{array}{l l}
1 - e^{\Pe \tilde{x}} & \quad   \tilde{x} < 0 \\
0  & \quad \tilde{x} > 0.
\end{array} \right.
\end{equation}
Finally, inserting the above into the front velocity condition \eqref{un2} we obtain the result $\hat{U}_0=1$, i.e. in the units chosen the planar reaction front advances with a unit velocity along the flow direction.

\section{Stationary dissolution fingers in 2D}
\label{sec:2d}

In this section, we derive the shape of a single dissolution finger propagating invariantly towards the undissolved matrix. We begin with the two-dimensional case.

At the boundary between the phases, Eq. \eqref{seq:bnd_eq} is satisfied. Without any loss in generality, one might extend this equation to the whole domain $\Omega$ as
\begin{equation}
F({\bf\hat r}, \hat t) \frac{\partial\hat c}{\partial\hat t} - |\hat\nabla\hat c|^2  = 0,
\end{equation}
with $F$ formally given as $F = |\hat\nabla\hat c|^2/\partial_{\hat t}\hat c$ and satisfying $\left. F \right|_{{\bf\hat r} \in \partial\Omega(t)} = \Pe$. To proceed, we use the ansatz due to Ivantsov \cite{Ivantsov1947,AddaBedia1994} by assuming that 
the unknown $F$ function dependence on the spatial and time coordinates is of the form
\begin{equation}
F({\bf\hat r}, \hat t) = F(c({\bf\hat r}, \hat t)),
\label{eq:ansatz}
\end{equation}
so that
\begin{equation}
F(\hat c) \frac{\partial\hat c}{\partial\hat t} - |\hat\nabla\hat c|^2  = 0 \label{eq:bnd_extended}
\end{equation}
is satisfied in the whole domain $\Omega$. Obviously, $F$ must satisfy
\begin{equation}
F(\hat c=0) = \Pe
\end{equation}
for the consistency with the boundary condition \eqref{seq:bnd_eq}. Using this idea, Ivantsov has found his famous set of steady-state solutions to the dendrite growth problem in a supercooled melt. Note that assuming the ansatz \eqref{eq:ansatz} might lead to losing some of more general solutions of the problem.

Let us now suppose that our problem has a stationary solution: a dissolution finger moving invariantly in the $x$ direction with velocity $U$. Rewriting Eq. \eqref{eq:bnd_extended} in the moving coordinate system gives
\begin{equation}
-\hat U F(\hat c) \frac{\partial\hat c}{\partial\hat x} - |\hat\nabla\hat c|^2 = 0
\end{equation}
or
\begin{equation}
\hat\nabla\hat c \cdot (\hat\nabla\hat c + \hat U F(\hat c) \hat\nabla \hat x) = 0.
\label{p5}
\end{equation}

Let us note that both the Laplace equation \eqref{b0} and the advection-diffusion equation with potential flow \eqref{b1} are conformally invariant. A large class of conformally invariant, non-Laplacian physical problems has been identified by Bazant in \cite{Bazant2003,Bazant2004} and involves such phenomena as nonlinear diffusion or advection or electromigration coupled to diffusion. The conformal invariance of these processes  provides an effective way of solving these problems \cite{Choi2005}. In the context of growth processes, these techniques have been used to track the evolution of the interface in solidification and melting under the action of a potential flow \cite{Maksimov1976,Goldstein1978,Kornev1994,Cummings1999,Cummings1999b,Bazant2003,Bazant2006}. However in these works the growth has been taking place in external flows (with the fluid outside of the growing object), in contrast to the present case where the medium is porous, and the driving flow goes through the growing finger and then into the undissolved medium.

In our case, Eq. \eqref{p5} is also written in a conformally invariant form. This allows us to make a conformal coordinate transformation $(\hat x,\hat y) \rightarrow (\xi,\eta)$ such that $\hat c=\hat c(\xi)$ (i.e., the isolines of $\hat c$ coincide with the curves $\xi(\hat x,\hat y)=\const$). Since Eq.~\eqref{p5} is conformally invariant, we get
\begin{equation}
\frac{d \hat c}{d \xi} + U F(\hat c(\xi)) \frac{\partial \hat x}{\partial \xi} = 0.
\label{p5b}
\end{equation}
The immediate conclusion is that $\frac{\partial\hat x}{\partial \xi}$ must be a function of $\xi$ only\footnote{Note that if one sought a stationary solution without having assumed the Ivantsov ansatz, one would obtain at this point
\begin{equation}
\frac{d \hat c}{d \xi} + U F(\xi,\eta) \frac{\partial \hat x}{\partial \xi} = 0
\end{equation}
which does not provide any information about the mapping.}, hence
\begin{equation}
\hat x(\xi,\eta) = \Phi_1(\xi) + \Phi_2(\eta).
\label{f1}
\end{equation}
The Cauchy-Riemann conditions,
\begin{equation}
\frac{\partial\hat y}{\partial \eta} = \frac{\partial\hat x}{\partial \xi}, \quad \frac{\partial\hat y}{\partial \xi} = -\frac{\partial\hat x}{\partial \eta},
\label{CR}
\end{equation}
lead to the Laplace equation for $\hat x$, $\nabla^2_{\xi\eta}\hat x = 0$, the general solution of which, compatible with \eqref{f1}, is
\begin{subequations}
\begin{equation}
\hat x(\xi,\eta) = \frac{1}{2} \hat{\rho} \left((\xi-\xi_*)^2 - (\eta-\eta_*)^2 \right).
\end{equation}
Using again \eqref{CR} allows us to find $\hat y(\xi,\eta)$ in the form
\begin{equation}
\hat y(\xi,\eta) = \hat{\rho} (\xi-\xi_*) (\eta-\eta_*) + \hat y_*.
\end{equation}
\label{eq:2d_map}
\end{subequations}
 Without the loss of generality we can set $\xi_* = \eta_* = y_* = 0$. Furthermore, we can assume that the parabola characterized by $\xi=1$ corresponds to the dissolution front (any other parabola can be scaled onto it by an appropriate choice of parameter $\hat{\rho}$). Note that the parameter $\hat{\rho}$ corresponds to the (dimensionless) radius of curvature of the parabolic finger. This suggests that the (dimensional) radius of curvature, $\rho$, could be used as a length parameter in our problem. Taking $l=\rho$ leads to $\hat{\rho}=1$ which simplifies the subsequent analysis.

Our final conclusion is the following: the only conformal mapping $(\hat x,\hat y) \rightarrow (\xi,\eta)$ which makes the concentration field in the dissolution finger a function of one variable only, $\hat c(\hat x,\hat y) = \hat c(\xi(\hat x,\hat y))$, is (up to translation)\footnote{In fact, we also assume the Ivantsov ansatz (Eq. \eqref{eq:bnd_extended}).}
\begin{align}
\hat x(\xi,\eta) &= \frac{1}{2} \left(\xi^2 - \eta^2 \right) \\
\hat y(\xi,\eta) &=  \xi \eta,
\end{align}
which is the usual transformation to the parabolic coordinates.

It is worth noting that parabolic geometries have also been obtained by other authors studying steadily translating growth forms. The classical example are Ivantsov's dendrites \cite{Ivantsov1947}. In fact, Adda Bedia and Ben Amar \cite{AddaBedia1994} have shown that, at least in the context of  steady state dendritic growth at zero surface tension, the parabolic solutions are the only admissible solutions once the Ivantsov ansatz is adopted and that even when one modifies this approach by making a more general ansatz, no new forms are found. Although the present case is of a more complicated nature, it is possible that also here the adoption of Ivantsov ansatz necessitates the appearance of the parabolic forms.

Once the shape of the finger is known, we can find the concentration and pressure both inside and outside of it. 

\subsection{Solution inside the finger}
\label{sec:finger_inside}

Inside the dissolution finger (i.e., for $\xi < 1$)
\begin{align}
& \Pe \kappa \frac{\partial \hat p}{\partial \xi}  \frac{d \hat c}{d \xi} +  \frac{d^2 \hat c}{d \xi^2}  = 0 \label{eq:c} \\
& \frac{\partial^2 \hat p}{\partial \xi^2} + \frac{\partial^2 \hat p}{\partial \eta^2} = 0 \label{eq:p2} \\
& \frac{d \hat c}{d \xi} + U F(\hat c(\xi)) \frac{\partial \hat x}{\partial \xi} = 0. \label{eq:F}
\end{align}

\begin{figure}
\centering
\includegraphics[width=.25\textwidth]{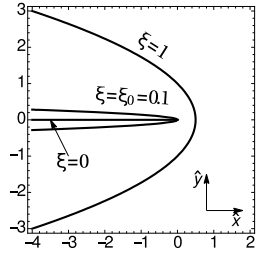}
\caption{Schematic representation of the boundary conditions for the concentration field $\hat c$ inside the parabolic dissolution finger. The finger is represented in the Cartesian coordinate system, with the $x$ axis pointing rightwards and the lengths scaled by the radius of curvature $\rho$. The outer parabola, denoted by $\xi=1$, constitutes the boundary of the finger and is characterized by vanishing of the reactants concentration, $\hat c(\xi=1) = 0$. At the inner parabola of $\xi=\xi_0$ (here $\xi_0 = 0.1$) the boundary condition \eqref{eq:vbc2} is imposed, $\hat c(\xi=\xi_0) = 1$. Additionally, the line $\xi=0$ is shown.}
\label{fig:parabolas}
\end{figure}

Additionally, at the front itself, the reactant concentration vanishes, i.e. $\hat c(\xi=1)=0$. It is harder to formulate the second boundary condition for the concentration, necessary to close the equation. As mentioned in \S\ref{sec:model}, the upstream condition in the form of Eq.~\eqref{eq:vbc2} is inconsistent with the presence of the finger extending towards $x \rightarrow -\infty$. Instead, we need to prescribe the concentration within the finger. In accordance with the assumption that 
$\hat c$ is a function of $\xi$ only, we impose $\hat c=1$ on one of the parabolas, defined by $\xi=\xi_0$ (see Fig. \ref{fig:parabolas}).

To proceed,  we conclude from Eq.~\eqref{eq:c} that $\partial_{\xi} \hat p$ must be a function of $\xi$ only. Along with the fact that $\hat p$ satisfies the Laplace equation \eqref{eq:p2} this gives, similarly as before,
\begin{equation}
\hat p(\xi,\eta) = -\frac{1}{2} \lambda (\xi^2-\eta^2) + A_1 \xi + A_2 \eta
\label{eq:p_ins}
\end{equation}
where the constant term has been dropped. However, the constants $A_1$ and $A_2$ have to be zero, otherwise the flow $\hat{\bf v} = -\kappa \hat \nabla \hat p$ would diverge at $\hat{\bf r}=0$. Inserting Eq. \eqref{eq:p_ins} back into \eqref{eq:c} yields for the concentration field
\begin{equation}
\hat c(\xi) = \frac{\text{erfi}(\alpha \xi) - \text{erfi}(\alpha)}{\text{erfi}(\alpha\xi_0) - \text{erfi}(\alpha)}, \label{eq:fing_c_sol}
\end{equation}
where $\alpha = \sqrt{\frac{1}{2}\lambda \kappa \Pe}$ and $\text{erfi}(x) \equiv \frac{2}{\sqrt{\pi}}\int_0^x e^{t^2} dt$.

Finally, Eq. \eqref{eq:F} allows us to find the velocity of the propagation of the finger
\begin{equation}
U  = \left.-\dfrac{c'(\xi)}{F(\hat c(\xi))\partial_\xi\hat x}\right|_{\xi=1} = \frac{2 \alpha\, e^{\alpha^2}}{\sqrt\pi\,  \,\Pe \left( \text{erfi}(\alpha) - \text{erfi}(\alpha\xi_0)\right)},
\end{equation}
where we have used the fact that  $F(\hat c(\xi=1)) = \Pe$.

\subsection{Solution outside the finger}

Outside the dissolution finger, in the undissolved matrix, the pressure still obeys the Laplace equation
\begin{equation}
\frac{\partial^2 \hat p}{\partial \xi^2} + \frac{\partial^2 \hat p}{\partial \eta^2} = 0. \label{v0}
\end{equation}
The continuity conditions for $\hat p$ and the $\hat v_n \equiv \hat v_\xi$ yield here
\begin{align}
-\frac{1}{2}\lambda (1-\eta^2) &= \left. \hat p \right|_{\xi\to 1^+} \label{v2} \\
-\kappa \lambda &= \left. \frac{\partial\hat p}{\partial\xi} \right|_{\xi\to 1^+}. \label{v1}
\end{align}
This is supplemented by the boundary condition \eqref{eq:vbc}
\begin{equation}
\left. -\hat \nabla \hat p \right|_{\xi\to\infty} = \hat {\bf e}_x = \hat \nabla \hat x,
\end{equation}
which can be written in the $(\xi, \eta)$ coordinates as:
\begin{equation}
\label{eq:bnd_downstream}
\lim\limits_{\xi\to\infty} \frac{1}{\xi}  \frac{\partial\hat p}{\partial \xi} = -1 \ \ \ \ \ \ \  
\lim\limits_{\xi\to\infty} \frac{1}{\eta} \frac{\partial\hat p}{\partial \eta} = 1. 
\end{equation}
The solution of Eq.~\eqref{v0} fulfilling (\ref{v2}--\ref{eq:bnd_downstream}) is
\begin{equation}
\hat p(\xi,\eta) = -\frac{1}{2}  (\xi^2-\eta^2) - (\kappa-1) (\xi-1).
\end{equation}
Additionally, the condition \eqref{v1} gives $\lambda=1$ for the constant appearing in the solution of the internal problem (Eqs. \ref{eq:p_ins}--\ref{eq:fing_c_sol}).

\subsection{Summary}

\begin{figure}
\centering
\includegraphics[width=0.9\textwidth]{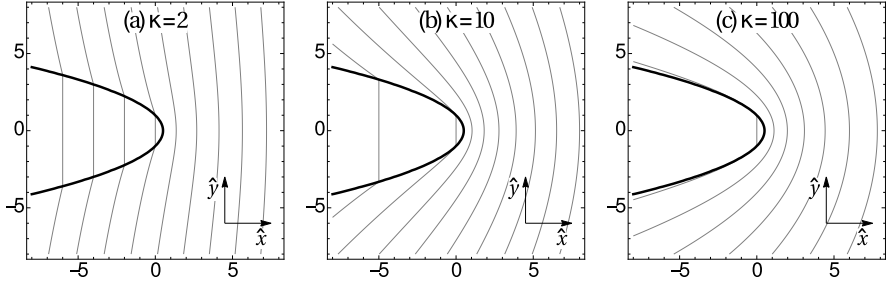}
\caption{Pressure field inside and outside of the two-dimensional stationary dissolution finger (marked by the thick parabola), for different values of permeability contrasts $\kappa$, $\kappa=2$~(a), $\kappa=10$~(b), and $\kappa=100$~(c). The (dimensionless) pressure drop between the neighboring isobars is 2~(a), 5~(b), and 50~(c).}
\label{fig:2D_pressure}
\end{figure}

The pressure field in the system is
\begin{equation}
\hat p(\xi,\eta) = \left\{
\begin{array}{l l}
-\frac{1}{2}  (\xi^2-\eta^2) & \quad \text{for $0 \leq \xi \leq 1$} \\
-\frac{1}{2}  (\xi^2-\eta^2) -(\kappa-1) (\xi-1) & \quad \text{for $\xi \geq 1$}.
\end{array}
\right.
\end{equation}
Fig. \ref{fig:2D_pressure} shows the pressure fields, plotted for various permeability contrasts $\kappa$, in the physical $(x,y)$ plane. Note that inside the finger the isobars are parallel to each other and oriented along the $y$ direction, which is a manifestation of the fact that the flow is uniform there:
\begin{align}
&\hat p(\hat x, \hat y) = -\hat x,  & \quad  {\bf r} \in \Omega, \\
&\hat {\bf v}(\hat x, \hat y) = \kappa {\bf \hat e}_x, & \quad {\bf r} \in \Omega.
\label{ins}
\end{align}
In the dimensional variables, the (constant) Darcy velocity inside the finger is
\begin{equation}
{\bf v}_{in}= \kappa v_0 {\bf \hat e}_x
\label{vin}
\end{equation}

Outside the finger, in the undissolved domain $\bar\Omega$, the pressure field is
\begin{equation}
\hat p(\hat x, \hat y) = -\hat x - (\kappa-1)\left( \sqrt{\hat r+\hat x} - 1 \right).
\label{out}
\end{equation}
The term linear in $\hat x$ dominates far from the boundary $\partial\Omega$, which agrees with the far-downstream condition for the flow field \eqref{eq:vbc}. Note that both the pressure field inside the finger \eqref{ins} as well as that outside \eqref{out} obey the upstream boundary condition \eqref{flo}.

The above-obtained pressure field is analogous to that derived by Kacimov and Obnosov~\cite{Kacimov2000}, who have been analyzing the groundwater flow in a porous medium with a parabolic inclusion. In fact the pressure problem solved in \cite{Kacimov2000} is more general than that considered here, since it involved the far-field flow velocity oriented at an arbitrary angle with respect to the parabola axis, not necessarily parallel as in \eqref{eq:vbc}. Interestingly, in each case the flow within the parabola is found to be uniform. This is the parallel of the classical result by Poisson and Maxwell who noted that the electric field inside elliptic or ellipsoidal inclusions is uniform \cite{Poisson1825,Maxwell1881}, see also Carslaw and Jaeger \cite{Carslaw1959} for the interpretation of this fact in the context of heat transfer.

Finally, let us note that the higher the permeability contrast $\kappa$, the more disturbed is the pressure field by the presence of the finger and the more aligned the isobars along the reaction front. In the limit of $\kappa \to \infty$, the dissolution front $\partial\Omega$ would corresponds to one of the isobars.

The concentration $\hat c$ is nonzero only inside the dissolution finger and depends on the $\xi$ coordinate in the following way,
\begin{equation}
\hat c(\xi) = \frac{\erfi\left(\sqrt{\frac{1}{2}\kappa\Pe}\xi\right) - \erfi\left(\sqrt{\frac{1}{2}\kappa\Pe}\right)}{\erfi\left(\sqrt{\frac{1}{2}\kappa\Pe}\xi_0\right) - \erfi\left(\sqrt{\frac{1}{2}\kappa\Pe}\right)}
\quad \text{for $\xi_0 \leq \xi \leq 1$}
\label{eq:2D_c}
\end{equation}
where, as mentioned above, the inlet concentration, $\hat c=1$, has been imposed on the parabola $\xi=\xi_0$ ($0 \leq \xi_0 < 1$). As a special case, one can choose $\xi_0 = 0$, which yields
\begin{equation}
\hat c(\xi) = 1-\frac{\erfi\left(\sqrt{\frac{1}{2}\kappa\Pe}\xi\right)}{\erfi\left(\sqrt{\frac{1}{2}\kappa\Pe}\right)}
\quad \text{for $0 \leq \xi \leq 1.$}
\end{equation}
The inspection of Eq.~\eqref{eq:2D_c} reveals that the relevant parameter controlling the concentration field  is the product 
\begin{equation}
\Pep = \kappa\Pe = \frac{\rho v_{in}}{\phi_1 D}.
\end{equation}
with $v_{in}$ given by Eq.~\eqref{vin}. This parameter is interpreted as a P\'{e}clet number related to the flow within the finger.

\begin{figure}
\centering
\includegraphics[width=0.9\textwidth]{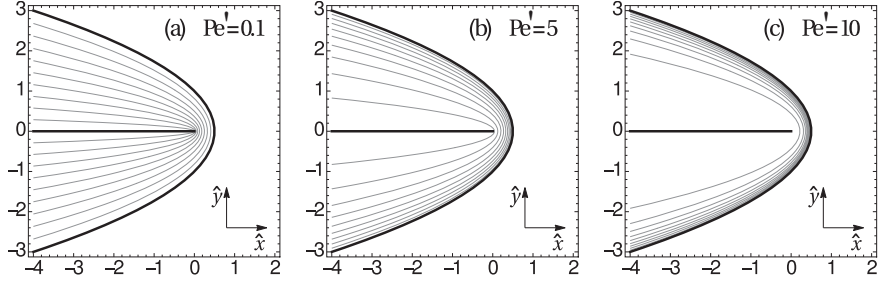}
\caption{Reactant concentration field $\hat c$ inside the two-dimensional stationary dissolution finger (marked by the thick parabola), for   $\Pep=0.1$ (a), $\Pep=5$ (b), and  $\Pep=10$ (c). The boundary condition for the concentration \eqref{eq:vbc2}, has been imposed on the line $\xi=\xi_0=0$. The (dimensionless) concentration drop between the neighboring isolines of $\hat c$ (thin parabolas) is set to $0.1$. The boundary of the finger is also an isoline of the concentration (for $\hat c=0$).}
\label{fig:2D_concentration}
\end{figure}

\begin{figure}
\centering
\includegraphics[width=.6\textwidth]{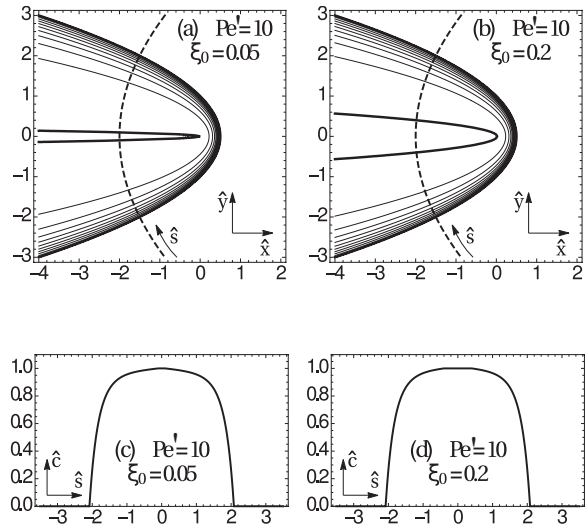}
\caption{(a,b) Reactant concentration field $\hat c$ inside the two-dimensional stationary dissolution finger in the ``convective'' regime, with  $\Pep=10$. The thick solid lines mark the finger boundary and the $\xi=\xi_0$ parabola with $\xi_0=0.05$ (a) and $\xi_0=0.2$ (b), on which the inlet concentration has been imposed, $\hat c(\xi=\xi_0) =1$. The (dimensionless) concentration drop between the neighboring isolines of $\hat c$ (thin parabolas) is set to $0.1$. The dashed line marks the parabola $\eta=2$ (locally orthogonal to the isolines of concentration) parametrized by the arc length $\hat s$. The reactant concentration profile along this parabola, $\hat c(\hat s)$, is plotted in panels (c) and (d).}
\label{fig:2D_concentration_varxi0}
\end{figure}

The impact of this parameter on the concentration profiles is analyzed in 
Fig. \ref{fig:2D_concentration}. For small values of $\Pep$, Eq. \eqref{eq:2D_c} is approximated by
\begin{equation}
\hat c \approx \dfrac{\xi-1}{\xi_0-1} \quad \text{for $\xi_0 \leq \xi \leq 1$}.
\end{equation}
i.e. the concentration depends linearly on the parabolic coordinate $\xi$. The gradient of the concentration field is then relatively uniform ({\it cf.} Fig. \ref{fig:2D_concentration}a). This situation can be referred to as the ``diffusive'' regime. In the opposite case of large $\Pep$ (the ``convective'' regime), the concentration boundary layer is formed, with uniform concentration $\hat{c}=1$ in the body of the finger and high gradients near the boundary (Fig. \ref{fig:2D_concentration}c). In this limit the solutions found by us  acquire direct physical relevance since a somewhat artificial choice of $\xi_0$ becomes irrelevant here, as long as the parabola $\xi=\xi_0$ lies in the bulk. The characteristic width of the boundary layer is $\left(\frac{1}{2}\Pep\right)^{-1/2}$, therefore the condition
\begin{equation}
\xi_0 \ll 1 - \left(\frac{1}{2}\Pep\right)^{-\frac{1}{2}}
\label{eq:xi0_constr}
\end{equation}
guarantees that $\xi_0$ lies in the bulk and thus its precise value does not affect the concentration profiles. This is confirmed by the comparison of $\hat{c}$ profiles for $\Pep=10$ and $\xi_0=0.05$ with those for $\xi=0.2$, which are found to be almost indistinguishable, as shown in Fig.~\ref{fig:2D_concentration_varxi0}. 


Finally, the advancement velocity of the dissolution finger is
\begin{equation}
\hat U  = \left. -\dfrac{\hat c'(\xi)}{F(\hat c(\xi)) \partial_\xi \hat x} \right|_{\xi=1} = \frac{\sqrt{2\kappa}\, \exp\left(\frac{1}{2}\Pep\right)}
{\sqrt{\pi\,\,\Pe} \left( \text{erfi}\left(\sqrt{\frac{1}{2}\Pep}\right) - \text{erfi}\left(\sqrt{\frac{1}{2}\Pep}\xi_0\right)\right)} \label{gre}
\end{equation}
with the two limiting cases (corresponding to small/high values of $\Pep$) given by
\begin{align}
& \lim_{\Pep\to 0} \Pe \hat U = (1-\xi_0)^{-1}, \\
& \lim_{\Pep\to\infty}\hat U = \kappa.
\end{align}
The velocity scale is $l/\tau = \gamma v_0$. Thus in the limit of small $\Pep$, the dimensional finger advancement speed approaches
\begin{equation}
U \approx \dfrac{\gamma v_0}{\Pe(1-\xi_0)} = \dfrac{\gamma \phi_1 D}{\rho(1-\xi_0)},
\label{smallPe}
\end{equation}
i.e. it is inversely proportional to the radius of curvature of the tip.  Such a behavior is in full analogy to the Ivantsov result for the dendrite propagation speed \cite{Ivantsov1947}. On the contrary, for large P\'eclet numbers the propagation velocity becomes constant with respect to the curvature of the finger tip and simply proportional to the Darcy velocity inside the finger
\begin{equation}
U \approx \gamma v_0 \kappa = \gamma v_{in}.
\end{equation}
Notice the inherent indeterminacy of the general result \eqref{gre}: we have obtained the relation between the tip radius and the propagation speed with no means of evaluating  each of these quantities independently. This is connected with the scale-invariance of the original equations (\ref{e1}--\ref{un}) and can be lifted only after the introduction of an additional lengthscale. Again, this feature is shared by both the Ivantsov parabolas and the Saffman-Taylor fingers, which also represent families of growth forms and need the short-scale regularization mechanism such as surface tension or kinetic undercooling for the selection of a particular tip radius and advancement velocity. In the case of dissolution patterns, the additional lengthscale might originate from the reaction front width (assumed to be zero in our analysis) or from the interactions with other dissolution fingers: in real system, the finger is never infinite, but always constrained by the presence of its neighbors. 

\section{Comparison with the Nilson \& Griffiths model}\label{Nilson}

At this point it is worth noting the main differences between our finger growth model and the one of Nilson and Griffiths \cite{Nilson1990}. First of all, we differ in the boundary conditions at infinity.  Whereas we assume that the entire system is flushed with a reactive fluid (with constant pressure gradient both at $x \rightarrow -\infty$ and $x \rightarrow \infty$), Nilson \& Griffiths assume constant pressure at $x \rightarrow \infty$, which means that the fluid far from the tip of the finger is kept immobile. Since at the same time the reactive fluid is constantly injected into the system, displacing the original pore fluid, the latter becomes compressed, which does not seem to be a realistic assumption under typical groundwater pressures. In another case considered in Ref.~\cite{Nilson1990} the medium is supposed to be initially unsaturated, with no fluid present. Then the reactive fluid is entering the system, creating a finger and flooding the medium in front of the finger tip, forming a 'product layer'. Due to the absence of imposed pressure gradient, the fluid in the product layer gradually slows down, as the thickness of the layer is increased until it reaches a stationary situation in which both finger and the product layer advance with equal speeds into the medium. Again, this is a fundamentally different situation from the one considered in the present work, where the entire medium is flushed with a reactive fluid due to the imposed pressure gradient. In particular, the solution tubes of Fig.~\ref{fig:solution_pipes} are formed under such conditions, with the gravity imposing the pressure gradient across the entire system. 

There is yet another difference between our approaches, this time related to the constraints on the values of the physical parameters, under which the finger model is supposed to work. Similarly to us, Nilson \& Griffiths adopt the thin-front approximation, stating that the reactant is entirely consumed as soon as it hits the finger boundary. As mentioned above, this corresponds to the assumption that $H=Dk/v_{in}^2 \gg 1$. At the same time, however, it is assumed in \cite{Nilson1990} that the concentration of the reactant inside the finger is constant, so that the dissolution front velocity is proportional to the local fluid velocity. This is only possible if the advection within the finger dominates over the diffusive transport, i.e. $\Pep=v_{in} l/D \gg 1$. The double limit (${Dk}/{v_{in}^2} \gg 1, \ \ \ v_{in}l/D \gg 1$) puts rather stringent constraints on the admissible values of $D$, $v_{in}$ and $k$ --- i.e. the flow rates should on one hand be large (to keep the transport advective within the finger), but on the other hand small enough to keep the reaction front thin. Conversely, in the present work, while keeping the front thin, we fully resolve the concentration field inside the finger, and thus obtain the finger propagation velocities over the entire range of P\'{e}clet numbers, {\it cf.} Eq.~\eqref{gre}. At smaller $\Pep$, the effects connected with the buildup of a diffusive layer of the depleted reactant appear \cite{Szymczak2013}, which have a significant impact on the wormhole propagation speed.

The differences in the boundary conditions between our system and the one of Ref.~\cite{Nilson1990} are reflected in different relations between the finger advancement velocity and its tip curvature. Nilson \& Griffiths find that the finger velocity is inversely proportional to its radius of curvature. We get a similar result,  but in the limit of small $\Pep$ only, when the diffusive effects play a central role in the dynamics. These effects are absent altogether in the Nilson \& Griffiths model. Conversely, we find that for large $\Pep$ the finger propagation velocity is independent of its curvature.

\section{Stationary dissolution fingers in 3D}
\label{sec:3d}

In this section we study the geometry of three-dimensional dissolution fingers. 
Again, we assume the Ivantsov ansatz as well as the stationarity of the fingers, so the starting point of our investigations is the following set of equations for pressure and concentration inside the finger    
\begin{subequations}
\label{eq:3d_statement}
\begin{align}
& \kappa \hat\nabla \hat p \cdot \hat\nabla \hat c + \Pe^{-1}\hat\nabla^2 \hat c = 0 \\
& \hat\nabla^2 \hat p = 0\\
& \hat\nabla \hat c \cdot ( F(\hat c)\,\hat U\, \hat \nabla \hat z +\hat\nabla \hat c) = 0 
\end{align}
\end{subequations}
where we have assumed the flow along the $z$ direction. As before, on the boundary of the finger ($\xi=1$) we put $\hat c = 0$ and $F=\Pe$. In the region outside the finger the pressure obeys the Laplace equation whereas the concentration vanishes.  

Let us switch to the cylindrical coordinate system, $(\hat x,\hat y,\hat z) \to (\hat r, \phi, \hat z)$ and look for the solutions with rotational invariance, i.e. independent of $\phi$. Eqs. \eqref{eq:3d_statement} in the new variables take the form
\begin{subequations}
\begin{align}
& \kappa \Pe \hat\nabla_{rz} \hat p \cdot \hat\nabla_{rz} \hat c + \hat\nabla_{rz}^2 \hat c + \frac{1}{\hat r} \frac{\partial\hat c}{\partial\hat r} = 0 &\quad \mathbf{r} \in \Omega(\hat t)  \label{seq:adv_diff} \\
& \hat\nabla_{rz}^2 \hat p + \frac{1}{\hat r} \frac{\partial\hat p}{\partial\hat r} = 0 &\quad \mathbf{r} \in \Omega(\hat t)  \label{seq:press_in} \\
& \hat\nabla_{rz} \hat c \cdot ( F(\hat c)\,\hat U\, \hat \nabla_{rz} \hat z +\hat\nabla_{rz} \hat c) = 0 &\quad \mathbf{r} \in \Omega(\hat t)  \label{seq:mapping} 
\end{align}
\end{subequations}
where $\hat \nabla_{rz} = \left(\partial_r, \partial_z  \right)$ acts as if the $(\hat r,\hat z)$ were Cartesian coordinates. Similarly, $\nabla_{rz}^2 = \partial_r^2 + \partial_z^2$.
Analogously to  the previous case we are looking for a conformal map $(\hat z, \hat r)\to(\xi,\eta)$ such that $\hat c$ becomes a function of one variable only, $\hat c = \hat c(\xi)$. Again, from Eq. \eqref{seq:mapping} we draw the conclusion that such a map need to be a parabolic one,
\begin{subequations}
\begin{align}
& \hat z = \frac{1}{2}(\xi^2 - \eta^2) \\
& \hat r =  \xi\eta,
\end{align}
\label{eq:3d_map}
\end{subequations}
with the radius of curvature of the finger chosen as the unit length. 
As before, we assume that the finger boundary coincides with the $\xi = 1$ paraboloid.

\begin{figure}
\centering
\includegraphics[width=0.9\textwidth]{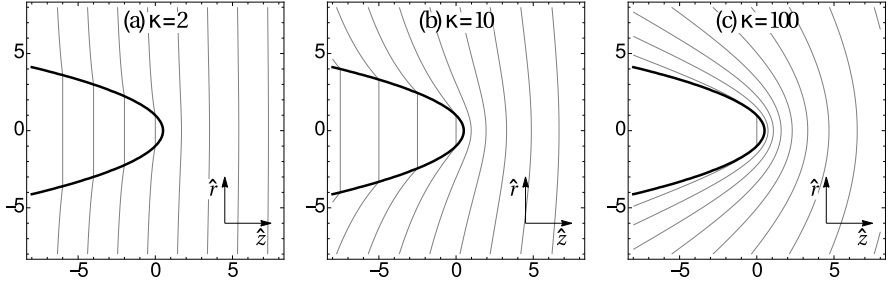}
\caption{Pressure field inside and outside of the three-dimensional stationary dissolution finger. The xz cross-section of the finger is shown, with the finger boundary marked by a thick parabola. The pressure field is plotted for different values of  permeability contrast: $\kappa=2$ (a),  $\kappa=10$ (b), and $\kappa=100$ (c). The (dimensionless) pressure drop between the neighboring isobars is 2 (a), 2.5 (b), and 10 (c).}
\label{fig:3D_pressure}
\end{figure}

Let us now use this mapping to solve the Laplace equation for the pressure inside and outside of the dissolution finger. Since the Lam\'{e} coefficients $L_\xi,L_\eta$ for the transformation \eqref{eq:3d_map} are equal to
\begin{equation}
L_{\xi} = L_{\eta} = \sqrt{\left(\frac{\partial\hat r}{\partial\xi}\right)^2 + \left(\frac{\partial\hat z}{\partial\xi}\right)^2} = \sqrt{\xi^2 + \eta^2},
\end{equation}
equation \eqref{seq:adv_diff} written in terms of $\xi$ and $\eta$ becomes
\begin{equation}
\Pep \frac{\partial\hat p}{\partial\xi} \hat c'(\xi) + \hat c''(\xi) + \frac{1}{\xi} \hat c'(\xi) = 0.
\label{eq:3d_c}
\end{equation}
Its form suggests that $\partial_{\xi}\hat p$ is a function of $\xi$ only and thus the pressure field $\hat p$ inside the dissolution finger is separable into
\begin{equation}
\hat p(\xi,\eta) = X(\xi) + Y(\eta).
\label{eq:3d_p_sep}
\end{equation}
The Laplace equation for the pressure inside the dissolution finger, Eq. \eqref{seq:press_in}, in the $\xi,\eta$ coordinates transforms into
\begin{equation}
\frac{\partial^2 \hat p}{\partial \xi^2} + \frac{\partial^2 \hat p}{\partial \eta^2} + \frac{1}{\xi} \frac{\partial\hat p}{\partial\xi} + \frac{1}{\eta} \frac{\partial\hat p}{\partial\eta} = 0, \quad 0 \leq \xi < 1.
\label{eq:3d_p}
\end{equation}
Inserting \eqref{eq:3d_p_sep} into Eq. \eqref{eq:3d_p} gives the result that the pressure field inside the dissolution finger should be of the form
\begin{equation}
\hat p(\xi,\eta) = -\frac{1}{2}\lambda_1 (\xi^2 - \eta^2) + C_1 \ln\xi + C_2\ln\eta + C_3, \quad 0 \leq \xi < 1.
\end{equation}
We set $C_1=C_2=0$, to avoid singularities at $\xi=0$ or $\eta=0$, while the constants $\lambda_1$ and $C_3$ are to be determined from the boundary and the continuity conditions. Next, we look for the solution of the Laplace equation for the pressure field outside of the finger of the same functional form as that inside,
\begin{equation}
\hat p(\xi,\eta) = -\frac{1}{2}\lambda_2 (\xi^2 - \eta^2) + D_1 \ln\xi + D_2\ln\eta, \quad \xi > 1.
\end{equation}
where the constant term has been skipped. 
The continuity conditions and the far-downstream boundary condition for the flow are analogous to the two-dimensional case (Eqs. (\ref{eq:vbc}--\ref{eq:pp})). The downstream boundary condition yields $\lambda_2=1$. The continuity of the pressure and the normal component of the flux at the finger boundary $\xi=1$ yields $D_2=C_3=0$, $\lambda_1=1$ and $D_1=1-\kappa$.
Eventually, the pressure field in the whole space is
\begin{equation}
\hat p(\xi,\eta) = \left\{
\begin{array}{l l}
-\frac{1}{2} (\xi^2-\eta^2) & \quad \text{for $0 \leq \xi \leq 1$} \\
-\frac{1}{2} (\xi^2-\eta^2) -(\kappa-1) \ln\xi & \quad \text{for $\xi \geq 1$}.
\end{array}
\label{flg}
\right.
\end{equation}
This solution has been plotted in Fig. \ref{fig:3D_pressure} for  different values of the permeability contrast, $\kappa$. The overall picture is similar to that in the two-dimensional case, with a uniform flow inside the finger and the flow disturbance around it relaxing to a uniform flow field far downstream.  
For small values of the permeability contrast $\kappa$, the presence of the finger does not significantly disturb the isobars, whereas in the opposite case the isobars become almost parallel to the finger boundary.

\begin{figure}
\centering
\includegraphics[width=0.9\textwidth]{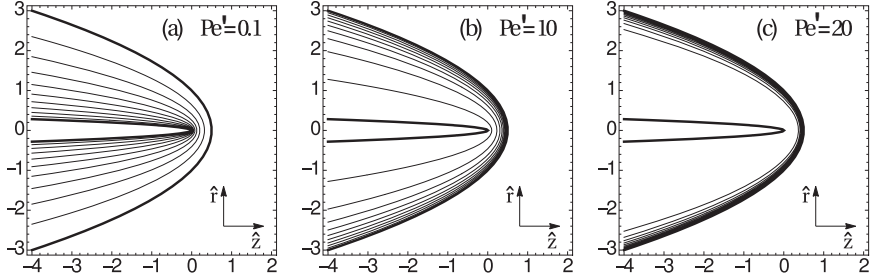}
\caption{Reactant concentration field $\hat c$ in the cross-section of the three-dimensional stationary dissolution finger, for $\Pep$=0.1 (a), $5$ (b), and $20$ (c). The parameter $\xi_0$ was chosen to be $0.1$. The (dimensionless) concentration drop between the neighboring isolines is set to $0.1$.}
\label{fig:3D_concentration}
\end{figure}

Using \eqref{flg} one can solve Eq. \eqref{eq:3d_c} and find the concentration field inside the finger. Formally, the solution can be written in terms of the exponential integral function $\Ei$\footnote{The exponential integral function is defined as $\Ei(x)\equiv - \int\limits_{-x}^{\infty}t^{-1}e^{-t} dt$.},
\begin{equation}
\hat c(\xi) = \frac{\Ei\left(\frac{1}{2}\Pep\xi^2\right) - \Ei\left(\frac{1}{2}\Pep\right)}{\Ei\left(\frac{1}{2}\Pep\xi_0^2\right) - \Ei\left(\frac{1}{2}\Pep\right)}, \quad \xi_0 < \xi < 1,
\label{eq:3d_conc}
\end{equation}
where, as before, we set the condition $\hat{c}=1$ on the parabola $\xi=\xi_0$ (with $0 < \xi_0 < 1$). The concentration fields for various parameters have been plotted in Fig. \ref{fig:3D_concentration}. Qualitatively, the behavior is similar to the one observed for the 2D case.  The ``diffusive'' regime is observed for small values of $\Pep$ (see Fig. \ref{fig:3D_concentration}a) and is characterized by relatively uniform distribution of the concentration gradient. In this case, the concentration field can be approximated by\footnote{$\Ei(x) = \gamma + \ln x + x + O(x^2)$}
\begin{equation}
\hat c(\xi) \approx \frac{\ln\xi}{\ln\xi_0}, \quad \xi_0 < \xi < 1.
\label{eq:3d_conc_small_args}
\end{equation}
In the opposite case of high $\Pep$ (see Fig. \ref{fig:3D_concentration}c), the boundary layer characterized by high gradients of the concentration $\hat c$ is formed near the finger boundary, while the in the interior of the finger the $\hat c$ field is almost uniform. The characteristic width of the boundary layer in the $(\xi,\eta)$-plane is, similarly as in the 2D case, $\left(\frac{1}{2}\Pep\right)^{-1/2}$, which sets the constraint on the choice of $\xi_0$, analogous to the two-dimensional one (Eq. \eqref{eq:xi0_constr}).
The only significant difference between the 3D and the 2D cases is that in 3D we cannot demand $\xi_0 = 0$, since the $\Ei$ function diverges at zero and the solution becomes singular. However, this constraint becomes irrelevant in the ``convective'' regime ($\Pep \gg 1$), since then $\hat c \approx 1$ throughout the bulk of the finger.

Finally, the finger propagation velocity, derived analogously to the 2D case,  is given by
\begin{equation}
\hat U = \left.- \dfrac{\hat c'(\xi)}{F(\hat c(\xi))\partial_\xi \hat z}\right|_{\xi=1} = \frac{2 \Pe^{-1} \exp\left(\frac{1}{2}\Pep\right)}{\Ei\left(\frac{1}{2}\Pep\right) - \Ei\left(\frac{1}{2}\Pep\xi_0^2\right)}.
\end{equation}
with the limiting cases characterized by
\begin{align}
\lim_{\Pep \to 0} \hat U \Pe &= - (\ln\xi_0)^{-1} \\
\lim_{\Pep \to \infty} \hat U  &= \kappa.
\end{align}
Again, there is a significant similarity with the 2D case. The (dimensional) propagation velocity of dissolution fingers at small P\'eclet numbers is inversely proportional to $\rho$,
\begin{equation}
U \approx -\dfrac{\gamma v_0}{\Pe\ln\xi_0} = -\dfrac{\gamma \phi_1 D}{\rho\ln\xi_0},
\end{equation}
whereas at high P\'eclet numbers the propagation velocity becomes independent of the radius of curvature and yields
\begin{equation}
U \approx \gamma v_{in}.
\label{eq:3d_U_highPe}
\end{equation}

\section{Summary and conclusions}
\label{sec:conclusions}

In this paper, we have been analyzing the geometry of the dissolution fingers, which are commonly found in soluble rocks. We have adopted the thin-front approximation, in which the reaction is assumed to take place instantaneously with the reactants fully consumed at the dissolution front. In this way the problem have become  Stefan-like with the undissolved phase downstream and the fully dissolved phase upstream. 
In such a setup we have posed the problem of existence and shape of steadily growing fingers. Two potentially limiting assumptions have been made. First, we have adopted the Ivantsov ansatz \eqref{eq:bnd_extended}, which has been successfully used in other studies of dendritic growth \cite{Ivantsov1947,Horvay1961,AddaBedia1994}, usually leading to the parabolic forms of advancing dendrites. Moreover, we have assumed that the transformation $(x,y) \rightarrow (\xi,\eta)$  from the original Cartesian coordinates to the coordinates spanned by the isolines and gradient lines of the reactant concentration field $c$ is conformal. Both of these assumptions limit the class of solutions that can be found with this technique, leading however to significant simplifications in the analysis, since the Ivantsov equation \eqref{p5} is 
conformally invariant and takes a particularly simple form \eqref{p5b} in the new coordinate system, where $c=c(\xi)$. The analysis of these equations leads to the conclusion that $(\xi,\eta)$ are parabolic coordinates and thus the advancing front is also of a parabolic form. The flow within the finger turns out to be uniform, in agreement with the earlier studies on the refraction of groundwater flow on parabolic inclusions \cite{Kacimov2000}. 
The magnitude of the flow is solely the function of the permeability ratio between the dissolved and undissolved matrix, $\kappa$. The concentration field is also relatively simple, and can be expressed in terms of the imaginary error function (in 2d, Eq. \ref{eq:2D_c}) or the exponential integral (in 3d, Eq.~\ref{eq:3d_conc}). This time, the parameter controlling the shape of the field lines is $\Pep$, defined as the ratio of convective to diffusive fluxes within the wormhole on the length scale of the radius of curvature of the finger tip. 

However, these simplifications come at a price, as we can only impose the concentration boundary conditions on the $\xi = \const$ lines, which limits the usefulness of these solutions in the physical applications. The most relevant physically seems to be the case of large P\'eclet numbers, since in that case the concentration within the finger becomes uniform with an exception of thin boundary layer at the reaction front itself and the exact placement of Dirichlet boundaries becomes irrelevant.

Finally, we have obtained the finger propagation velocity, which turns out to be inversely proportional to the radius of curvature in the small $\Pep$ limit and constant for large $\Pep$. Interestingly, the former prediction seems to be confirmed by the analysis of the shapes of the solution pipes --- the inspection of Fig.~\ref{fig:solution_pipes} and other photographs from this site reveals that the longer fingers have invariably smaller radii of curvatures.

The scale invariance of the problem leads to the dynamical indeterminacy: any advancement velocity is permissible provided that the radius of curvature of the finger is appropriately tuned, there is no selection principle to fix the tip curvature and the finger propagation velocity independently. Finding such a selection principle is in general a highly nontrivial task \cite{Pelce2004} and involves the introduction of additional length scale into the system. In the case of viscous fingering and dendritic growth this is achieved through the short-scale regularization mechanisms such as surface tension or kinetic undercooling. It is not clear what would be the physical origin of such a short-scale regularization in the present case. A candidate for the additional length scale could be the reaction front width, which in our analysis is zero, but remains finite in any physical case. Another possibility is that the additional length scale sought here is connected with the presence of other fingers, as it is always the case in natural systems ({\it cf.} Fig. \ref{fig:solution_pipes}). The full solution of the problem would then require matching the solution near the parabolic tip with that in the neighborhood of the root of the finger, where it joins with its companion.

\section*{Acknowledgments}

This work was supported by the National Science Centre (Poland) under research Grant No.
2012/07/E/ST3/01734. Pawe{\l} Kondratiuk is a beneficiary of the project ``Scholarships for PhD students of Podlaskie Voivodeship'' co-financed by the European Social Fund, the Polish Government and Podlaskie Voivodeship. The authors benefited from discussions with Tony Ladd and Krzysztof Mizerski.

\bibliographystyle{siam}

\end{document}